\documentclass[twocolumn,showpacs,preprintnumbers,amsmath,amssymb,superscriptaddress,10pt,aps,prb]{revtex4-1}
\usepackage{epsfig,amsopn}
\usepackage{graphicx}
\usepackage{epstopdf}
\usepackage{sidecap}
\usepackage{xcolor}
\usepackage{amsmath,amssymb}
\usepackage{amsthm}
\usepackage{enumerate}
\begin{document}
\title{Transport through Andreev bound states in a Weyl semimetal quantum dot}
\author{Dibya Kanti Mukherjee}
\affiliation{Harish-Chandra Research Institute, Chhatnag Road, Jhunsi, Allahabad 211019, India.}
%\author{Dibya Kanti Mukherjee}
%\affiliation{Harish-Chandra Research Institute, Chhatnag Road, Jhunsi, Allahabad 211 019, India.}
%\affiliation{Department of Physics, Indiana University, Bloomington, IN 47405}
\author{Sumathi Rao}
\affiliation{Harish-Chandra Research Institute, Chhatnag Road, Jhunsi, Allahabad 211019, India.}
\author{Arijit Kundu}
\affiliation{Physics Department, Indian Institute of Technology Kanpur, Kanpur 208016, India.}

\begin{abstract}
We study transport through a Weyl semimetal  quantum dot sandwiched between an $s$-wave superconductor and a normal lead. 
The conductance peaks at regular intervals and exhibits  double periodicity  with respect to  two characteristic frequencies of the system, one that originates from 
Klein tunneling in the system and the other coming from the chiral nature of the excitations. Using a scattering matrix 
approach as well as a lattice simulation, we demonstrate the universal features of the conductance through the system
and discuss the feasibility of observing them in experiments.
\end{abstract} 
\pacs{74.45.+c, 74.50.+r, 73.21.-b}

\maketitle
\emph{Introduction.}---%
Weyl semimetals (WSMs)~\cite{Vishwanath2011,Burkov2011a,Burkov2011b,Zyuzin2012a,Hosur2012} are 3D topological systems with an even number of \textit{Weyl nodes} in the bulk, with low-energy excitations having a definite chirality when the Fermi energy is near the Weyl nodes. The study of such systems has exploded in recent times both in the 
theoretical~\cite{Vazifeh2013,Son2013,Turner2013,Biswas2013,Hosur2013,Burkov2014,Gorbar2014,Uchida2014, Khanna2014,Ominato2014,Sbierski2014,Burkov2015a,Burkov2015b,Goswami2015,Baum2015, Khanna2016,Behrends2016,Rao2016,Baireuther2016a,Tao2016,Marra2016,Li2016,Baireuther2016b,Madsen2016,Obrien2017,Khanna2017,Bovenzi2017} as well as the  experimental~\cite{Xu2015a,Xu2015b,Lv2015a,Lv2015b,Lu2015,Jia2016} front. The reason for this  excitement is the non-trivial physics that can arise in Weyl systems, such as broken chiral symmetry (or the chiral anomaly) and Fermi arcs where the Fermi vector at one  surface is a discontinuous arc  that  connects to the other surface through the bulk, giving rise to exotic physical properties and transport signatures. 

\begin{figure}[t]
\centering
\includegraphics[width=0.48\textwidth]{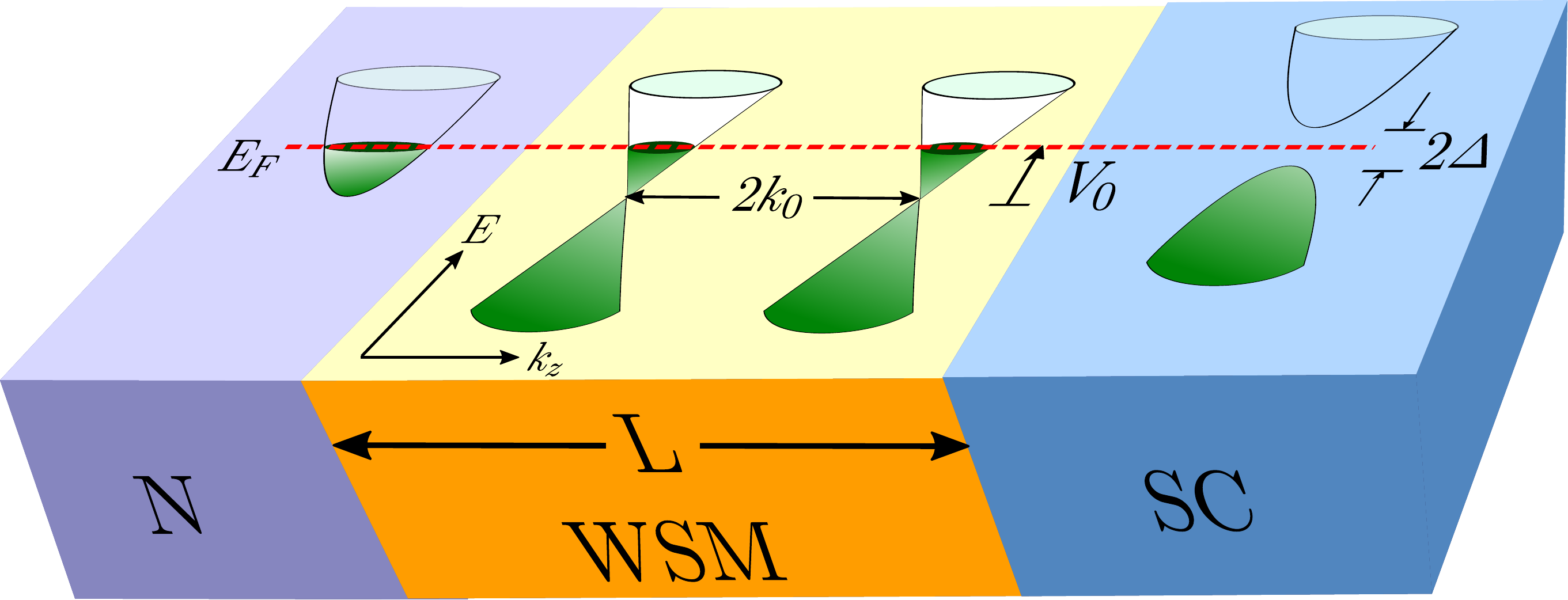} 
\caption{Setup of the system. A time-reversal broken Weyl semimetal WSM of length $L$ has been sandwiched between a superconductor (SC) with a gap $\Delta$ and  a normal/WSM metal lead (N). The momentum separation between the Weyl nodes in the WSM dot is $2k_0$ and the WSM has a bias $V_0$. }\label{fig:setup}
\end{figure}

A quantum dot made  of WSM material in the presence of superconductors  is of particular interest due to the distinctive  nature of transport at a WSM-Superconductor (SC) interface~\cite{Uchida2014,Khanna2016,Obrien2017,Khanna2017,Bovenzi2017} and provides the possibility of capturing the otherwise elusive  physics associated with the chiral excitation in the WSM~\cite{Obrien2017}. In this manuscript we study transport through the Andreev states of a WSM quantum dot in a simple setup where we sandwich the dot in between a superconductor and a normal lead (see Fig.~\ref{fig:setup}). Bound levels will form in the dot due to multiple reflections from the two boundaries  and these levels will strongly depend on the Fermi-energy mismatch \cite{Beenakker2006} between the dot and the SC, as well as on the size of the dot.  
One expects some of the physics of a graphene quantum dot~\cite{graphenedot} to carry over to this case, since the WSMs also have a linear dispersion; however there are differences as well. One of the
features of the Dirac dispersion is that the Andreev bound states  carry current that oscillates as a function of  $\chi = V_0 L/v_F$,  where $V_0$ is the chemical potential of the dot~\cite{Bhattacharjee2006,Bhattacharjee2007},
$L$ is its size and $v_F$ is the Fermi velocity. A second oscillation appears as a function of $\delta k L $, where 
$\delta k$ is the momentum separation of the nodes that are connected by superconducting pairing. In graphene,  an $s$-wave superconductor couples electrons at one valley with  holes at the other valley and the
Andreev bound states are hence also dependent on the matching of the valley polarizations~\cite{Akhmerov2007}, with $\delta k=K-K'$ as the separation of the valleys in  momentum space. On the other hand in a WSM-SC interface, the $s$-wave superconductor is required to  couple the  electrons at one node with the  holes  at the other node. Hence,  reflection processes couple one chiral node to another node of opposite chirality~\cite{Uchida2014,Khanna2016} and $\delta k = 2k_0$ where $2k_0$ is the distance between the nodes in momentum space. Coupling between nodes is  otherwise forbidden, irrespective of their positions in momentum space. Further, the inter-valley length scale $K-K'$ in graphene is quite large, whereas in WSMs,  $2k_0$ is a relevant length scale, because the nodes are typically quite close to each other. At finite bias, however, as we shall see below, the relevant parameter changes from $2k_0 L$, and the nature of the bands becomes important. In the rest of this paper, our focus is to study and predict the behavior  of   the current through the  Andreev bound  states of the WSM  quantum dot at a finite bias.

%------------------------------
%------------------------------

The central result of our work is to disentangle the periodicity of the conduction peaks of the WSM quantum dot (with a superconductor on one edge) due to the chirality of the nodes from the periodicity due  to the finite bias and Klein tunneling.  We find that at finite bias the conduction peaks follow a periodic pattern of the form $(q_+ \pm q_-)L \approx 2n_{\pm}\pi$, with $n_{\pm}$ being integers.  Here $q_{\pm}$ are the Fermi momenta in the quantum dot at finite bias,
 along the direction of conduction, and are the analogs of $k_0$ at zero bias.  Their values can be determined from the band structure of the system and the bias $V_0$ present in the dot. At small enough bias, the periodicity reduces to the expected $2k_0L = n\pi$ oscillations~\cite{Khanna2016,Khanna2017}.

\emph{Model and setup.}---%
The simplest model of a WSM with broken time-reversal (TR) symmetry requires two chiral nodes in momentum space, whereas the simplest WSM with broken inversion symmetry requires the presence of four chiral nodes. In the main text we restrict ourselves to using the simplest model of a TR-broken WSM, having two nodes, for analytic simplicity.
%Although this, in principle, fails to point out certain aspect of the physics we discuss, but is much easier to work with in analytical framework. 
We consider an inversion symmetry broken model, which also has some new aspects beyond what is present in the two node model, in the appendix.

A two-band TR-broken WSM model can be obtained by starting  from a four-band Hamiltonian describing a 3D TI in the Bi$_2$Se$_3$ family and including a time-reversal breaking perturbation $b_z$~\cite{Vazifeh2013} - 
%regularized on a simple cubic lattice (where we take the lattice spacing to be the unit of length) and we add a 
\begin{align}\label{eq:ham}
  H_{0} =& \epsilon_k \tau_x - \lambda_z \sin k_z \tau_y  \nonumber \\
  & -\lambda \tau_z \left( \sigma_x\sin k_y  - \sigma_y \sin k_x\right) + b_z \sigma_z + V_0.
\end{align}
where $\epsilon_k = \epsilon - 2{\tilde t} \sum_i \cos k_i$ is the kinetic energy and $\lambda, \lambda_z$ are spin-orbit coupling strengths. In the limit $\lambda_z \ll \epsilon -6{\tilde t} \ll b_z$, \cite{Khanna2014}, this gives a WSM phase, where the gap closes at momentum points $(0,0,\pm k_0)$, where ${\tilde t}k_0^2 \approx b_z - \epsilon + 6{\tilde t}$. A gate potential $V_0$ is applied to the dot region which spans a distance $L$. For sufficiently small $V_0$, the low energy excitations can be described by the  two-band Hamiltonian
\begin{align}\label{eq:HWSM}
H_{\text{WSM}} = \tilde{\epsilon}_k\sigma_z + \lambda (k_x\sigma_x + k_y\sigma_y) + V_0,
\end{align}
with $\tilde{\epsilon}_k \approx {\tilde t}(p^2+k_z^2-k_0^2)$  and with $p^2 = k_x^2+k_y^2$. 
In the rest of the paper, all parameters are scaled with respect to 
 ${\tilde t}$ which is the  energy  scale. The eigenvalues of Eq.~(\ref{eq:HWSM}) are $E_{\pm}(\mathbf{k})=\pm\sqrt{{\tilde \epsilon}_k^2 + \lambda^2p^2} + V_0$. This implies that the Fermi velocity is anisotropic -  the velocity in the  $z$ direction is different from that 
in the  $x,y$ direction. Close to the Weyl nodes $k_z=\pm k_0$, the Fermi velocity along the $z$-direction $v_z = 2k_0$. 

%------------------------------
\begin{figure}[t]
\centering
\includegraphics[width=0.45\textwidth]{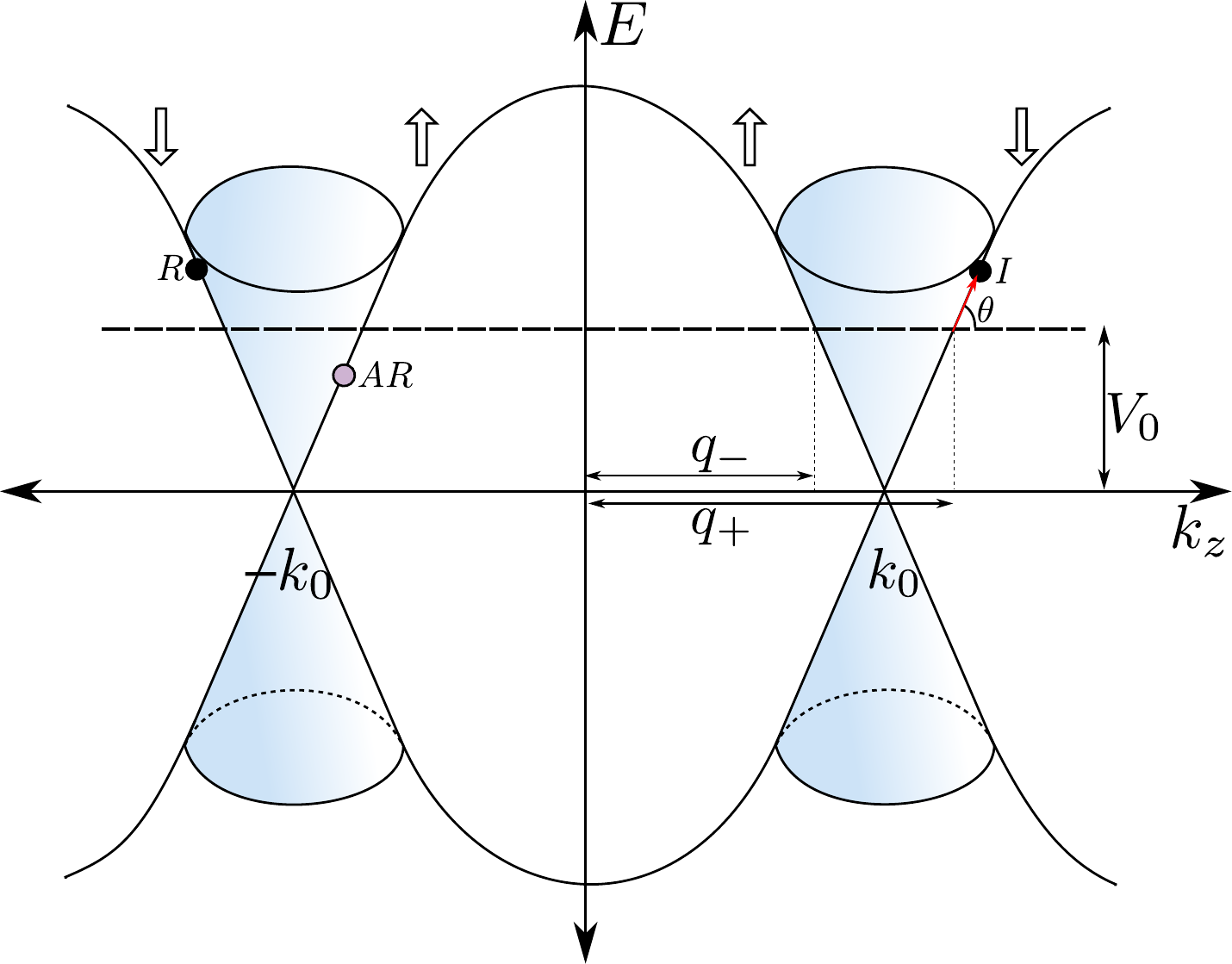} 
\caption{Diagrammatic representation of the possible scattering processes showing all the relevant scales. The Weyl nodes are located at $k_z=\pm k_0$ and $q_{\pm}$ are the two possible momenta of electronic excitations above the finite potential barrier $V_0$ of the WSM. tan$\theta$ denotes the Fermi velocity $v_F$ of such excitations. $I$ describes an incident electron and $R$ and $AR$ describe normal and Andreev reflected electrons and holes respectively. }\label{fig:nodes}
\end{figure}
%------------------------------

We construct a WSM dot by sandwiching the dot region (with a finite $V_0$) in between a normal-metal (N) and an $s$-wave superconductor (S). We then study  transport through the quantum dot, first using   a scattering matrix approach,  where
the $N$ region is chosen to be an unbiased WSM ($V_0 = 0$) and we use Eq.~(\ref{eq:HWSM}) to  solve for  the wavefunctions.  Next, we further study and verify our findings using a lattice simulation where we model the normal metal using a  \textit{flat band} approximation, i.e, by considering a uniform  density of states within the relevant energy scales.

The superconducting region can be described in terms of the Boguliobov-de Gennes (BdG) Hamiltonian: 
\begin{align}\label{eq:SC}
 H_{\text{SC}} = \left(\begin{array}{cc} \xi_k I_{2\times 2}& \Delta i  \sigma_y\\
-\Delta i \sigma_y  & -\xi_k I_{2\times 2}\end{array} \right), 
\end{align}
where $\Delta$ is the pairing potential in the superconductor and $\xi_k = (\hbar^2 (k_x^2+k_y^2+k_z^2)/2m_S - \mu_S)$. $m_S$ is the effective mass of the electron in the superconductor (we take $m_S \approx m_W$ for simplicity) and $\mu_S$ is the chemical potential. The parameter $\mu_S$ depends on the details of the superconducting material. In the numerical results shown, we take  $\mu_S \gg \Delta$, which is the  realistic limit.

\emph{Scattering matrix approach.}---%
Using familiar methods of solving for the wavefunction and matching them at the two boundaries, we obtain the net reflection matrix of the form 
\begin{align}
\mathcal{S}(E, \mathbf{p}) =\left(\begin{array}{cc}
r_{ee}(E, \mathbf{p}) & r_{he}(E, \mathbf{p}) \\
r_{eh}(E, \mathbf{p}) & r_{hh}(E, \mathbf{p})
\end{array}\right),
\end{align}
where, $r_{ee}$ and $r_{hh}$ are the reflection matrices, and  $r_{eh}$ and $r_{he}$ are the Andreev reflection matrices, in the basis of excitations near the two nodes with $\pm$ chirality~\cite{Uchida2014,Khanna2016}. $E$ is the incident energy and $\mathbf{p} = (p_x,p_y,0)$ is the momentum in the transverse direction. The differential conductance is then written as
\begin{align}
G_{\mathbf{p}}(E) =& \frac{e^2}{h}Tr[I_{2}-R_{ee}(E, \mathbf{p})R_{ee}(E, \mathbf{p})^{\dagger} \nonumber\\
 & \quad \quad + R_{he}(E, \mathbf{p})R_{he}(E, \mathbf{p})^{\dagger}] \label{condp}
\end{align}
where,
\begin{align}
R_{ee(he)}=\left(\begin{array}{cc}
               \sqrt{v_{e(h)}^+} & 0 \\
               0 & \sqrt{v_{e(h)}^-}
             \end{array}\right)
                   r_{ee(he)}
                   \left(\begin{array}{cc}
               \frac{1}{\sqrt{v_e^+}} & 0 \\
               0 & \frac{1}{\sqrt{v_e^-}}
            \end{array}\right) \nonumber
\end{align}
where $v_{e(h)}^j$ is the velocity of the electron (hole) channel of the $j$th node. The nature of processes at the WSM-SC boundary is depicted in Fig.~\ref{fig:nodes}. The relation in Eq. \ref{condp} is true for each momentum $\mathbf{p}$in the transverse direction. Finally, we integrate over the transverse momentum to obtain the current $G(E) =\sum_{\textbf{p}} G_{\mathbf{p}}(E)$.

%The zero bias conductance of this geometry is plotted with the length of the \textcolor{blue}{insulating region} sandwiched between the normal and superconducting leads. From fig. [?] it is clear that this involves two frequencies: one which has its origin in the height of the potential barrier in the \textcolor{blue}{insulating region}, the other due to the chiral nature of the nodes that always forces the electrons to be reflected from one node to the other as either an electron or a hole.  

%------------------------------
%\begin{figure}[t]
%\centering
%\includegraphics[width=0.48\textwidth]{Num.pdf} 
%\caption{The schematic of the procedure used for lattice simulation. After integrating out the two leads, one superconducting and one normal metal, the full Greens function of system, $\mathcal{G}$, contains the corresponding self energies. The final current through the system is obtained after averaging over the lead states, which provides the information of the Fermi function of the leads.}\label{fig:num}
%\end{figure}
%------------------------------

%------------------------------
\begin{figure}[t]
\centering
\includegraphics[width=0.43\textwidth]{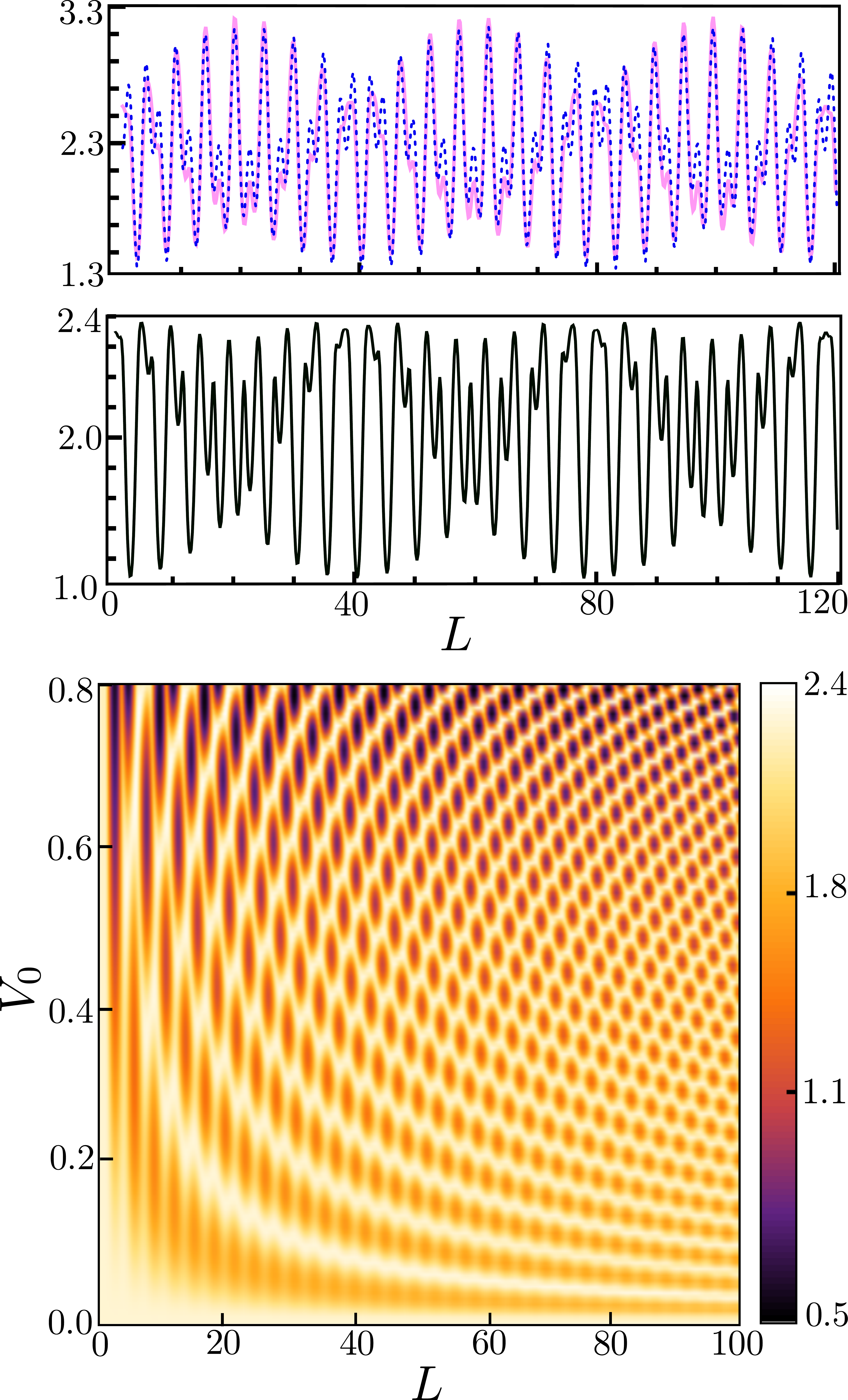} 
\caption{(color online) A typical pattern of the current through the WSM dot oscillating with the size of the dot, with beats due to the double periodicity (a) 
The peaks of conduction, i.e, the map of the Andreev spectrum, appear at lengths $d$ where $(q_+ + q_-)d/\pi$ is an integer, with $q_{\pm} = \sqrt{k_0^2\pm 2m_WV_0}$ in red(solid) lines. The best fit to this pattern in terms of  the simple two frequency function given in Eq. \ref{eq:func}  with $\alpha=2.24$ and $\beta =0.9$ plotted in blue (dotted) lines is also shown. Note the excellent agreement between the numerical data and the formula.  The parameters used are $k_0=1$, $m_S=m_W=0.5$, $\mu_S=4$, $\mu_W=0$, $\lambda=0.5$, $\Delta=0.01$, $V_0=0.56$. (Here we only consider normal incidence). (b) 
 Here the current integrated over the transverse momentum is shown, which  also  peaks at the same lengths $d$ where
 $(q_+ + q_-)d/\pi$ is an integer.
 (c) The complete plot of the current though the WSM dot as a function of its bias voltage and its size. Other parameters used are mentioned above.}\label{fig:scat1}
\end{figure}
%------------------------------

%------------------------------
\begin{figure*}

\centering
\includegraphics[width=1.0\textwidth]{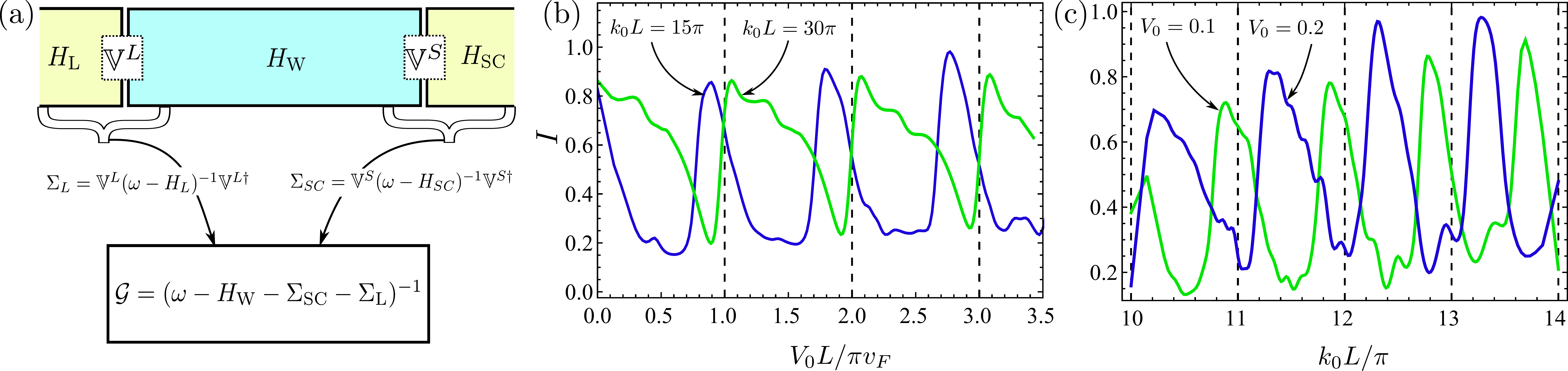} 
\caption{(color online) (a)The schematic of the procedure used for the lattice simulation. After integrating out the two leads, one superconducting and one normal metal, the full Green's function of the system, $\mathcal{G}$, contains the corresponding self energies. The final current through the system is obtained after averaging over the lead states, which include the information of the Fermi function of the leads (b), (c) Results of the lattice based simulation, verifying the oscillation dependence  of the current  as a function of $V_0$ and $k_0$ respectively. Here, the dotted lines show the periodicity expected from Eq. \ref{eq:func2}.
The parameters used are $\Delta=0.1$, $\epsilon=6$, $\lambda_Z=\lambda=0.5$, $\mu_L=0.05$, $\mu_R = 0$. The length of the WSM dot is kept fixed at 60 in units of lattice spacing.} \label{fig:num}
\end{figure*}
%------------------------------

We summarize our results from the scattering matrix approach in Fig.~\ref{fig:scat1} and we emphasize the following: first, the oscillation in the conductance is present even for normal-incidence, as expected from earlier results~\cite{Uchida2014,Khanna2016, Bovenzi2017}  which showed that the probability of normal-reflection at a WSM-SC junction is finite at normal incidence. Next, the oscillations in the  conductance appear due to multiple reflections in the dot region, similar to those  of a quantum mechanical double barrier problem. But for a WSM, such reflections can only take place from one chiral node to  the other chiral node of opposite chirality (c.f. Fig.~\ref{fig:nodes}), with inter-nodal distance $2k_0$. At finite bias, due to the presence of $V_0$, the relevant length scale depends on a combined function of $k_0$ and $V_0$,
i.e.,  they depend on $q_{\pm} =\sqrt{k_0^2\pm 2m_W V}$ , which are
momenta along the direction of propagation at the Fermi energy in  the dot-region.  This allows us to predict the oscillation frequencies depending on the symmetry, the positions of the Weyl nodes, the bias, etc.  In the present model, the conductance can be fitted well with the functional dependence of the form
\begin{align}\label{eq:func}
G = \alpha + \beta \sin\left[(q_+ + q_-)L\right] \sin\left[(q_+ - q_-)L\right],
\end{align}
where, $\alpha, \beta$ are independent of the length $L$, and  can, in principle, be obtained analytically, as shown in the appendix.  In Fig.~\ref{fig:scat1}(a),  we  show the pattern of the conductance obtained at \textit{normal incidence},  $G_0$,  fitted with a function of the form given in Eq.~(\ref{eq:func}). The close correspondence shows that the theoretically obtained function can predict all the peaks in the conductance $G$.  In Fig.~\ref{fig:scat1}(b), we show the full conductance, after integrating over the transverse momenta. The conductance continues to peak at values of $L$ where $(q_+ + q_-)L/\pi$ is an integer. Finally,  in Fig.~\ref{fig:scat1}(c) we show the variation of $G_0$ as functions of both the barrier height $V_0$ and $k_0$.  This  pattern can be fully predicted from the functional dependence in Eq.~(\ref{eq:func}).

Note that  for $V_0 \ll k_0^2$, $q_{\pm} \approx k_0 \pm (m_W V_0/k_0)$. We also note that the amplitude of the velocity at the Fermi energy in the dot-region is $v_F = k_0/m_W$. So, the conductance oscillations  have a slow frequency 
envelope whose period is $ V_0L/v_F =n\pi$ and a faster oscillation characterized by $k_0 L = m\pi$, (where $n,m$ are integers), allowing us to write the conductance as
\begin{align}\label{eq:func2}
G \approx \alpha + \beta \sin\left(2k_0L\right) \sin\left(2V_0L/v_F\right),
\end{align}
with corrections to the above equation appearing only at the order $\mathcal{O}\left(V_0^2/k_0^2\right)$.
Note however, that in Fig.~\ref{fig:scat1}, 
we have specifically chosen a value of $V_0$, such that condition for Eq.~(\ref{eq:func2}) is not satisfied.
In the regime, where the condition for Eq.~(\ref{eq:func2}) is satisfied, we find that  the periodicity for the conductance shows
peaks as a function of $L$ and $V_0$ whenever $k_0L = n\pi$ and $V_0L/v_F = n\pi$ as expected.

Finally, we also note that the amplitude $\beta$  of conductance oscillations depends strongly on the ratio $k_F/k_0$ and increases with increasing $V_0$. On the other hand, $\beta$ decreases with increasing incident energy $E$ and the conductance reaches a maximum value of  $4e^2/h$, and becomes independent of the barrier height $V_0$ in the limit $E \rightarrow \Delta$, matching earlier results in similar systems like graphene~\cite{Beenakker2006,Bhattacharjee2006}. We discuss the dependence of $\beta$ on $E$ and other parameters in the appendix.  In passing, we also note that a similar functional dependence (as shown in Eq.~(\ref{eq:func})), of the conductance oscillations would be true for a graphene dot, when  $2k_0$ and $v_F$ are respectively replaced by the momentum separation between the two valleys of graphene $K-K'$ and the Fermi velocity near the Fermi energy. 

%------------------------------
%\begin{figure}[t]
%\centering
%\includegraphics[width=0.45\textwidth]{v0k0.pdf} 
%\caption{The results of lattice based simulation, verifying the oscillation conditions of current stepulating in Eq.~\ref{eq:func}. (a) Shows the oscillation with $V_0$. The Fermi velocity, $v_F$ depends strongly on $V_0$ and deviates from the value $2k_0$ at large $V_0$, that we extract from the band structure of the system, Eq.~\ref{eq:HWSM}. The parameters used are $t=1$,$\Delta=0.1t$,$\epsilon=6t$,$\lambda_Z=\lambda=0.5t$ In (b), we show the second oscillation with $k_0$.}\label{fig:k0v0}
%\end{figure}
%------------------------------

\emph{Lattice simulation.}---%
In order to study  transport in our geometry, we implement a slight modification of  the standard Landauer-Buttiker formalism to suit our purpose.  We write the Fourier transformed Hamiltonian of Eq.~(1) and include a normal lead and a superconducting lead on the two  sides of the system along the z-axis, with tunneling matrices $\mathbb{V}^S$  and $\mathbb{V}^L$ respectively as shown in  Fig.\ref{fig:num}(a) (see appendix C for details). By integrating out the lead degrees of freedom, we obtain the Green's function for the whole system $\mathcal{G}$ as
\begin{align}
  \mathcal{G}^{-1}(\omega) = \mathcal{G}_W^{-1}(\omega)-\Sigma_{SC}(\omega)-\Sigma_{L}(\omega),
\end{align}
where $\mathcal{G}_W$ is the Green's function for the isolated WSM dot$, \mathcal{G}_W^{-1}(\omega)_{ij} = \omega\delta_{i,j} - {H_W}_{ij}$; $\Sigma_{SC}$ and $\Sigma_L$ are, respectively, the self energies due to the superconducting and normal leads, expressed as $\Sigma_{SC}(\omega) = \mathbb{V}^{S}\mathcal{G}_{SC}(\omega)\mathbb{V}^{L\dagger}$ and $\Sigma_{L}(\omega) = \mathbb{V}^{L}\mathcal{G}_{L}(\omega)\mathbb{V}^{L\dagger}$. Here $\mathcal{G}_{SC}$ and $\mathcal{G}_L$ denote the Green's function of the isolated superconducting and normal leads. Further, we implement a $flatband$ approximation for the Green's function of the normal lead,
where the density of states of the lead, $\rho_L (\omega)$ is taken to be a constant independent of the energy, and so
$\mathcal{G}_{L} = -i\pi \rho_L$. % and in this limit, $\Sigma_L = -i\pi \mathbb{B}^{\dagger}\rho_L I_{\tau}\otimes I_{\sigma}\mathbb{B}$.
The Green's function for the superconducting lead  $\mathcal{G}_{SC}$ is obtained by recursively solving for the surface Green's function of the $s$-wave superconductor~\cite{Lopez1985}. A schematic diagram that represents this process is presented in Fig.\ref{fig:num}(a).

We then compute the current flowing from a site $z$ to $z+1$ in the WSM dot given by
\begin{align}
  J_z(t)&=-\frac{2e}{\hbar}(-{\tilde t}+\lambda_z\tau)\text{Im}\langle \Psi^{\dagger}_{z+1,\bar{\tau},\sigma}(t)
\Psi_{z,\tau,\sigma}(t)\rangle
\end{align}
where $\Psi_{i,\tau,\sigma}$ is a column matrix representing the annihilation operator at site $i$ in the basis of orbital index $\tau = - {\bar{\tau}} \pm$ and spin index $\sigma=\pm$.  
%This is a direct application of the  Landauer-Buttiker formalism, but differs in particulars  from methods utilized in studying systems with superconducting leads\cite{}. 
The information about  the chemical potential of the leads (and the temperature, in principle) is included when averaging over the lead states. We show in the appendix C that this current can be written in terms of the Green's function of the full system $\mathcal{G}(\omega)$, at zero temperature, as
\begin{align}
J_z = \sum_{P = \text{SC,L}}e\text{Im}\int d\omega \text{Tr}\Big[ \mathcal{A}\mathcal{G}_{z,I}(\omega)\zeta^P(\omega)\mathcal{G}^{\dagger}_{z+1,I}(\omega)\Big]
\end{align}
where $\mathcal{A}_{31} = \mathcal{A}_{42} = -{\tilde t}+\lambda_z$ and $\mathcal{A}_{13} = \mathcal{A}_{24} = -{\tilde t}-\lambda_z$ and $\mathcal{A}_{ij} = 0$ otherwise. Here, $\zeta^{SC}(\omega) = \mathbb{V}\rho_{\text{SC}}(\omega)\mathbb{V}^{S\dagger}$ and  $\zeta^L(\omega) = \mathbb{V}^L\rho_{\text{L}}(\omega)\mathbb{V}^{L\dagger}$. We further consider the simplest case when the system is translation invariant in the transverse direction, so that the  transverse momentum is just a parameter.

We obtain the current as a function of $k_0$ and $V_0$ with the chemical potential on the left lead kept fixed at $\Delta/2$, and summarize the results in Figs.~\ref{fig:num}(b) and (c), where we have also taken the transverse momentum to be zero. As in the scattering matrix calculation, here again, the current oscillates as a function of both $k_0L/\pi$ and $V_0L/\pi v_F$, which clearly confirms the central result of our paper that inter-node Andreev reflection, if not prohibited by additional symmetries of the problem~\cite{Bovenzi2017}, plays a crucial role in determining transport properties of the Weyl semimetal-superconducting interface.

The distinct unambiguous signatures of WSM systems can be further clarified  if one takes an inversion symmetry broken WSM. An inversion broken WSM requires the presence of at-least four chiral nodes in the Brillouin zone. In the simplest situation, the nodes can be co-linear in momentum space, and without loss of generality, can be placed at momentum $\mathbf{k}_1=(-k^+,0,0), ~\mathbf{k}_2=(-k^-,0,0), ~\mathbf{k}_3=(k^-,0,0),  ~\mathbf{k}_4=(k^+,0,0)$. Time reversal symmetry requires the first and last nodes to have the same chirality, and the two nodes in the middle to  have opposite chirality.  If the chirality of the nodes were not relevant -i.e., if we were working with a 3 dimensional Dirac metal, then
%If nodes would not have definite chirality, then a $s$-wave proximity would couple, through Andreev processes, nodes 1-4 and 2-3, giving the relevant momentum change $2k^{\pm}$. Whereas, in a WSM the coupling is only allowed between nodes 1-2, and 3-4, giving relevant momentum $k^+-k^-$ and between nodes 1-3, and 2-4, giving relevant momentum $k^++2k^-$. Such a situation can strongly distinguish a Dirac metal from WSM, but working with 4-band model is cumbersome in a scattering matrix framework. We discuss such  model in Supplemental Material.
proximity to an s-wave superconductor  would couple  nodes of opposite momenta through Andreev processes.
So we would expect the relevant momentum scales to be  $2k^{\pm}$. But  in a WSM the coupling is 
only allowed between nodes 1-2, and 3-4, giving the relevant momentum scale  $k^+-k^-$ and between nodes 1-3, and 2-4, giving the relevant momentum scale $k^++k^-$.  Thus  the relevant scales of the conductance oscillations strongly distinguishes between a dot made of a Dirac metal from a dot made of a WSM. However, working with  a 4-band model is cumbersome in the  scattering matrix framework. We discuss the lattice results of such a WSM dot in the appendix.

\emph{Summary.}---%
To summarise, we have discussed transport through a Weyl semimetal quantum dot, in a normal-metal-WSM-superconductor geometry, that captures a number of features unique to the presence of chiral nodes in the WSM. We took a simple time-reversal broken WSM and studied it in the  scattering matrix approach as well as by using  tight-binding simulations. The key result of our work, Eq.~(\ref{eq:func}), differentiates the effect of Klein tunneling in the Dirac system from that due to the presence of chiral nodes in the WSM. An experimental setup should be similar in essence to that shown in  Ref.~\onlinecite{graphenedot}, but the details of the prediction would depend on the material used.

%---------------------------------------------------

\section*{Appendix}

\subsection{Solving the scattering problem}
In this section, we describe the derivation of the scattering matrix in a Normal-WSM dot-SC system. As described in the main text, the normal Hamiltonian is modelled by a WSM Hamiltonian without any chemical potential whereas the WSM dot is modelled by the same WSM Hamiltonian along with a barrier potential $V_0$. We define $V(z)=V_0(\Theta(z)-\Theta(z-L))$ where we assign the locations of the Normal-WSM dot junction and the WSM dot-SC junctions to be at $z=0$ and $z=L$ respectively. The wavefunction corresponding to energy $E$ in the normal system(for $z<L$) is given by the following energy eigenstates of Eq.~(2) of the main text in the Nambu-Gor'kov space (with the Hamiltonian in the hole space written as $-H_{\text{WSM}}^*(-\mathbf{k})$),
\begin{align}
\psi_{\text{N}}(z<0) =\sum_{\sigma=\pm}\Bigg\{& \mathcal{E}^{\sigma} \left(a^{\sigma}_Re^{\sigma ik_e^{\sigma}z} +a^{\sigma}_Le^{-\sigma ik_e^{\sigma}z} \right) \nonumber \\
 +&\mathcal{H}^{\sigma} \left(b^{\sigma}_Re^{-\sigma ik_h^{\sigma}z} +b^{\sigma}_Le^{\sigma ik_h^{\sigma}z} \right)\Bigg\},
\end{align}
and similarly, the wavefunction in the WSM dot corresponding to the same energy is given by:
\begin{align}
\psi_{\text{WSM}}(0<z<L) =\sum_{\sigma=\pm}\Bigg\{& \mathcal{E}^{\sigma} \left(c^{\sigma}_Re^{\sigma ik_e^{\sigma}z} +c^{\sigma}_Le^{-\sigma ik_e^{\sigma}z} \right) \nonumber \\
 +&\mathcal{H}^{\sigma} \left(d^{\sigma}_Re^{-\sigma ik_h^{\sigma}z} +d^{\sigma}_Le^{\sigma ik_h^{\sigma}z} \right)\Bigg\}.
\end{align}
Here $\sigma = \pm$ is the band index, $a_i,c_i(b_i,d_i)$ denote the electron (hole) amplitudes with $i\in\{L,R\}$ denoting the left or right moving solution.  $\mathcal{E}^{\sigma}(\mathcal{H}^{\sigma}$) are normalized eigenvectors, which are non-zero in electron (hole) sector of the Hamiltonian. 
In each sector $\mathcal{E}(\mathcal{H})^+\propto(f_{e(h)},(-)\lambda_{+(-)})^T$, and  $\mathcal{E}(\mathcal{H})^-\propto((-)\lambda_{-(+)},f_{e(h)})^T$, with $f_{e(h)} = \mu_W + V(x) + (-)E_{i}+\sqrt{(\mu_W + V(x) +(-)E_i)^2 - (\lambda p)^2}$, $\lambda_{\pm} = \lambda(k_x + i k_y)$. 

In the superconductor, the solutions of Eq.~(3) of the main text are: 
\begin{align}
\psi_{\text{SC}}(z>L)=\left(\begin{array}{c} uc_{\uparrow}\\ uc_{\downarrow}\\ -vc_{\downarrow}\\ vc_{\uparrow}\end{array}\right) e^{ i q_ez} +\left(\begin{array}{c} vd_{\downarrow}\\ -vd_{\uparrow}\\ ud_{\uparrow}\\ ud_{\downarrow}\end{array}\right) e^{ -i q_hz}, \nonumber
\end{align}
where, with $\Omega = \sqrt{\Delta^2-E_i^2}$,
$$ u(v) = \sqrt{\left(E_i +(-)i\Omega\right)/2E_i}$$
and $q_e$ and $-q_h$ are, respectively, the outgoing electron and hole momenta in the superconductor, defined as (with Fermi momentum $k_F$)
\begin{align}
 q_{e(h)} = \sqrt{k_F^2-p^2+(-)2m_S i \Omega/\hbar^2} ~.\nonumber
\end{align}
The boundary conditions at $z=\{0,L\}$ are given by the continuity of the wavefunction and its derivative at that point:
\begin{align}
& \psi_{\text{N}}(z)=\psi_{\text{WSM}}(z)|_{z=0} \nonumber \\
& \psi_{\text{WSM}}(z)=\psi_{\text{SC}}(z)|_{z=L} \nonumber \\
& \partial_z\psi_{\text{WSM}}(z)\mid_{z=0} = \partial_z\psi_{\text{N}}(z)\mid_{z=0} \nonumber \\
& m_S\left(\begin{array}{cc} \sigma_z & 0\\ 0 & \sigma_z\end{array}\right)\partial_z\psi_{\text{WSM}}(z) = m_W\partial_z\psi_{\text{SC}}(z)\mid_{z=L},\nonumber 
\end{align}
with $\sigma_z$ being the Pauli matrix. As was mentioned in the main text, we take $m_S \approx m_W$ for simplicity. By solving these equations, we get the reflection matrices,
\begin{align}\label{eq:Smat}
 &\left(\begin{array}{c}
        a_L^+\\
       a_L^-\\
        b_L^+\\
        b_L^-\\
       \end{array}
\right) = \left(\begin{array}{cc}
                 r_{ee} & r_{eh} \\
                 r_{he} & r_{hh} \\
                \end{array}
 \right) \left(\begin{array}{c}
        a_R^+\\
        a_R^-\\
        b_R^+\\
       b_R^-\\
       \end{array}
\right)
\end{align}
which were used in the main text.

%------------------------------
\begin{figure}[ht]
\begin{center}
\includegraphics[width=0.48\textwidth]{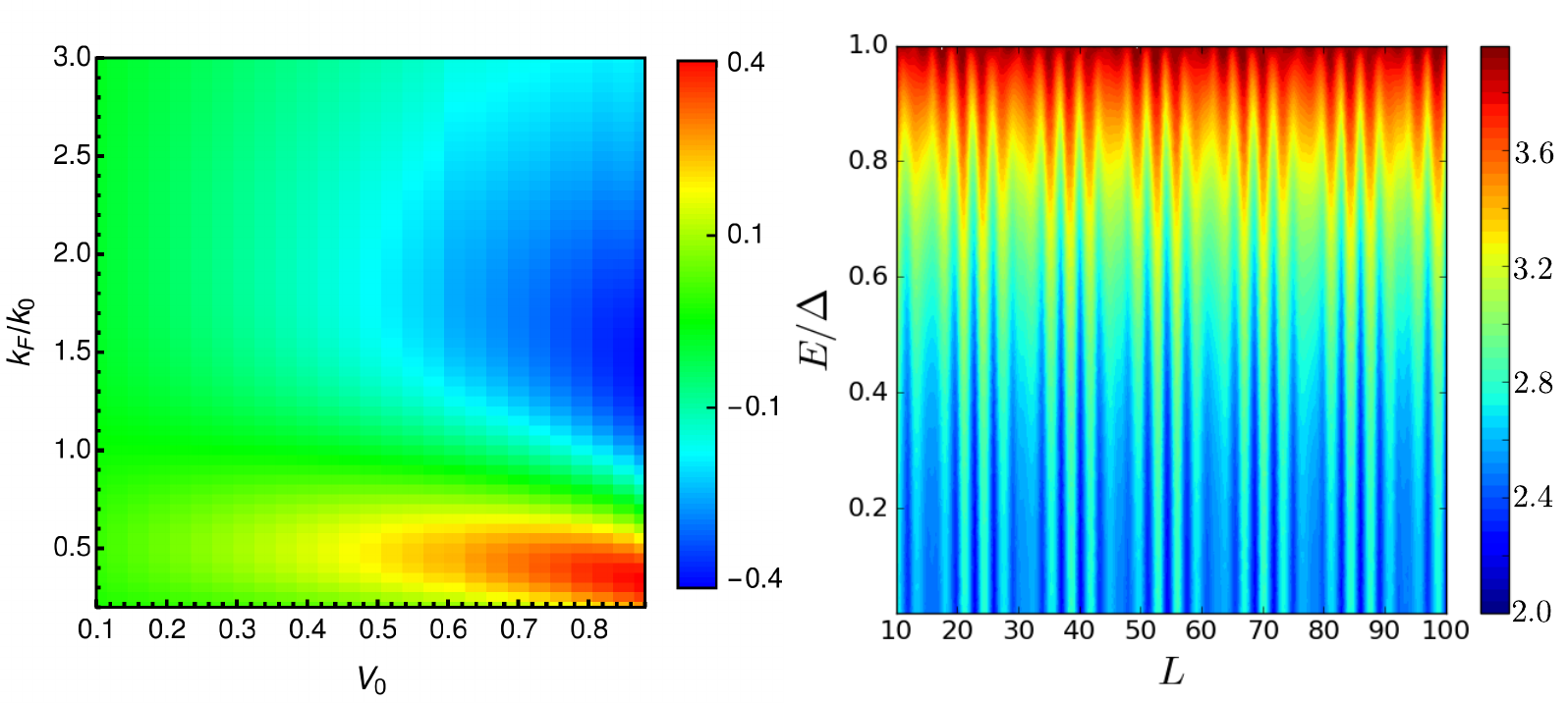} 
\end{center}
\caption{(a) Variation of the amplitude of oscillation ($\beta$) of the zero-bias conductance with the barrier height along  the $x$ axis and the ratio of the Fermi momentum and the separation of Weyl nodes along the $y$ axis. The parameters used are $k_0=1$, $m_S=m_W=0.5$, $\lambda=0.5$, $\Delta=0.01$. (Here we only consider normal incidence). (b) Conductance as a function of the length of the barrier along the $x$ axis and the incident energy along the $y$ axis at fixed $V_0=0.2$}\label{fig:b}
\end{figure}
%------------------------------

\subsection{Variation of the parameter $\beta$  with the system parameters  and the conductance with incident energy}
The amplitude of oscillation $\beta$ (see Eq.(6) of the main text) depends strongly on the system parameters, especially on the position of the Fermi vector $k_F$ of the superconductor and generally increases with increasing $V_0$ due to the Fermi energy mismatch. We show the numerical fitting of $\beta$ in the phase space of $k_F-V_0$ in Fig.~\ref{fig:b}(a). 

With increasing incident energy $E$, the net conductance reaches a universal value of $4e^2/h$ when $E/\Delta$ reaches unity  as depicted in Fig.~\ref{fig:b}(b).

\subsection{Details of the tight-binding simulation}
Here we briefly describe how we arrive at Eq.~(10) of the main text. For the TR symmetry broken Weyl semi-metal, the Hamiltonian is written as $H^0 = H_C+H_{SO}+H_E$, where,
\begin{align}
  H_C =& -{\tilde t}\sum\limits_{\langle\textbf{r},\textbf{r}'\rangle}\psi_{\textbf{r}}^\dagger\eta^z\tau^xI_\sigma\psi_{\textbf{r}'} + \epsilon\sum\limits_{\textbf{r}}\psi_{\textbf{r}}^\dagger\eta^z\tau^xI_\sigma\psi_{\textbf{r}'} + h.c. \nonumber \\
  H_{SO} =& i\lambda \sum\limits_{\textbf{r}}(\psi_{\textbf{r}}^\dagger\eta^z\tau^z\sigma^y\psi_{\textbf{r}+x}+\psi_{\textbf{r}}^\dagger\eta^z\tau^z\sigma^x\psi_{\textbf{r}+y} )  \nonumber \\
       & + i\lambda_z\sum\limits_{\textbf{r}}\psi_{\textbf{r}}^\dagger\eta^z\tau^yI_\sigma\psi_{\textbf{r}+z} + h.c. \nonumber \\
  H_E =& \sum\limits_{\textbf{r}}\psi^\dagger_{\textbf{r}}( b_0\eta^z\tau^y\sigma^z  -b_xI_\eta\tau^x\sigma^x+b_yI_\eta\tau^x\sigma^y \nonumber  \\ & + b_zI_\eta I_\tau\sigma^z)\psi_{\textbf{r}}~.
\end{align}
Here $\psi_{i,\sigma,\eta}^{\dagger}$ is the creation operator of electron with spin $\sigma$ ($=\uparrow,\downarrow$) and with orbital index $\eta$ ($=1,2$) at site $i$ of the WSM.  We consider the $x,y$ directions to be  translationally  invariant, so that the momenta  $k_x,k_y$  appear as parameters. After Fourier transforming in the $x,y$ directions, our next step is to rewrite the Hamiltonian in the Nambu-Gorkov form - 
\begin{align}
  H^W &= \frac{1}{2}\sum\limits_{\langle z,z'\rangle}\Psi^{\dagger}_{z,i}h_{W ij}(k_x,k_y)\Psi_{z',j},
\end{align}
using  the basis 
\begin{align}
    \Psi^{\dagger}_z = \big(&\psi^{\dagger}_{z,\uparrow,1},\psi^{\dagger}_{z,\downarrow,1},\psi^{\dagger}_{z,\uparrow,2},\psi^{\dagger}_{z,\downarrow,2}, \nonumber \\
& \psi_{z,\downarrow,1},-\psi_{z,\uparrow,1},\psi_{z,\downarrow,2},-\psi_{z,\uparrow,2}\big) \nonumber.
\end{align}
For each site $z$, the basis $\Psi_{z,i}$ has 8 components for $i=1,..,8$. The superconductor is modeled as a 1D superconductor: 
\begin{align}
    H^S= & \sum\limits_{z}\Phi_{z}^\dagger(\epsilon_{SC}\eta^z+\Delta\eta^x)I_\sigma\Phi_{z} \nonumber \\
    & -t_{SC}\sum\limits_{\langle z,z'\rangle}\Phi_{z}^\dagger\eta^z I_\sigma\Phi_{z'}+hc, \nonumber \\
    \equiv &\frac{1}{2}‎‎\sum\limits_{\langle z,z'\rangle}\Phi^{\dagger}_{z,i} h_{Sij}\Phi_{z',j}.
\end{align}
where $\Phi^{\dagger}_z = \big(\phi^{\dagger}_{z,\uparrow},\phi^{\dagger}_{z,\downarrow},\phi_{z,\downarrow},-\phi_{z,\uparrow}\big)$. The normal lead's Hamiltonian is the written as:
\begin{align}
  H^L &= ‎‎\frac{1}{2}\sum\limits_{\langle z,z'\rangle}a^{\dagger}_{z,i} h_{Lij}a_{z',j},
\end{align}
in the basis $a^{\dagger}_z =  \big(\alpha^{\dagger}_{z,\uparrow},\alpha^{\dagger}_{z,\downarrow},\alpha_{z,\downarrow},-\alpha_{z,\uparrow}\big)$.

The tunneling Hamiltonian between the WSM and the superconductor and  between the WSM and the normal leads are given respectively by: 
\begin{align}
  H^{WS}&=\frac{1}{2}\Psi^{\dagger}_{N,i}\mathbb{V}^S_{ij}\Phi_{1,j} + \frac{1}{2}\Phi^{\dagger	}_{1,i}\mathbb{V}^{S\dagger}_{ij}\Psi_{N,j}, \nonumber 
  \\ {\rm and}~
  H^{WL}&= \frac{1}{2}\Psi^{\dagger}_{1,i}\mathbb{V}^L_{ij}a_{N,j} + \frac{1}{2}a^{\dagger}_{N,i}\mathbb{V}^{L\dagger}_{ij}\Psi_{1,j}~.
\end{align}
Here, $\phi^{\dagger}$ and $a^{\dagger}$ are, respectively, the creation operators at the superconductor and the normal lead, without any orbital index. Also note that  we couple both orbitals equally to the superconducting site, which, albeit not the most generic case, represents the simplest coupling.

With this choice of basis, 
\[
\mathbb{V}^{i =SC/L}= 
  \begin{pmatrix}
    {t^{i}} & 0 & {t}^{i} & 0 & 0 & 0 & 0 & 0 \\
    0 & {t}^{i} & 0 & {t}^{i} & 0 & 0 & 0 & 0 \\
    0 & 0 & 0 & 0 & -{t}^{i} & 0 & -{t}^{i} & 0 \\
    0 & 0 & 0 & 0 & 0 & -{t}^{i} & 0 & -{t}^{i}
  \end{pmatrix}^T,
\] 
where $t^i = t^{SC/L}$ are the hopping matrix elements between the leads and the WSM.
The Hamiltonian has  an explicit particle-hole symmetry under
\begin{equation}
  \Phi^{\dagger}_{z,i} = \mathbb{C}_{ij} \Phi_{z,j}, ~~a^{\dagger}_{z,i} = \mathbb{C}_{ij} a_{z,j},  ~~\Psi^{\dagger}_{z,i} = \mathbb{C}^W_{ij} \Psi_{z,j}
\end{equation}
where, $\mathbb{C} = \sigma^y\otimes \sigma^y$, and, $\mathbb{C}^W = \sigma^y\otimes\mathbb{I}\otimes\sigma^y$.

Now, we wish to compute how the field operators evolve in time. Starting from the Heisenberg equation of motion
\begin{equation}
  \dot{a}_{z,i} = \frac{i}{\hbar}\big[H^L + H^{WL}, a_{z,i}\big],
\end{equation}
we obtain
\begin{align}
    &\dot{a}_z = \frac{i}{\hbar}(-h_L a_z-\mathbb{V}^{L\dagger}\Psi_1\delta_{z,N}) \nonumber \\
    &\Rightarrow\Big(i\hbar\frac{\partial}{\partial t}-h_L\Big)a_z=\mathbb{V}^{L\dagger}\Psi_1\delta_{z,N}.
\end{align}
The solution for the operator is given by
  \begin{align}
    a(t) &= i\hbar\mathcal{G}_L(t-t_0)a(t_0) + \int_{t_0}^t dt' \mathcal{G}_L(t-t')\mathbb{V}^{L\dagger}\Psi(t')\nonumber\\
    &= \eta_L(t) + \int_{t_0}^t dt' \mathcal{G}_L(t-t')\mathbb{V}^{L\dagger}\Psi(t'),
  \end{align}
where the Green's function $\mathcal{G}_L$ of the uncoupled lead is the solution of the equation 
\begin{equation}
  \big(i\hbar\frac{\partial}{\partial t}-h_L\big)\mathcal{G}_L(t-t') = \mathbb{I}\delta(t-t').
\end{equation}
Similarly, for the superconducting lead, one obtains
\begin{align}
    \Phi(t) &= i\hbar\mathcal{G}_S(t-t_0)\Phi(t_0) + \int_{t_0}^t dt' \mathcal{G}_S(t-t')\mathbb{V}^{S\dagger}\Psi(t')\nonumber\\
    &= \eta_S(t)+ \int_{t_0}^t dt' \mathcal{G}_S(t-t')\mathbb{V}^{S\dagger}\Psi(t')
\end{align}
Finally, for the operators in the Weyl semi-metal, we write:
\begin{equation}
  \dot{\Psi} = \frac{i}{\hbar}\big(-h_W\Psi-\mathbb{V}^La-\mathbb{V}^S\Phi\big).
\end{equation}
In the above equation, we need to substitute the solutions of $a(t)$ and $\Phi(t)$. We define the self energy operators as
  \begin{align}
    &\Sigma_L(t)=\int_{t_0}^t dt' \mathbb{V}^L\mathcal{G}_L(t-t')\mathbb{V}^{L\dagger} \nonumber \\ {\rm and}~
    & \Sigma_S(t)=\int_{t_0}^t dt' \mathbb{V}^S\mathcal{G}_S(t-t')\mathbb{V}^{S\dagger}.
  \end{align}
Fourier transforming the equation for $\Psi(t)$, we obtain 
\begin{equation}
  \Psi(\omega)=\mathcal{G_W}(\omega)\Gamma(\omega) \nonumber
\end{equation}
where $\mathcal{G_W}=(\omega-h_W/\hbar-\Sigma_L(\omega)/\hbar-\Sigma_S(\omega)/\hbar)^{-1}$ is the Green's function of the whole system and  $\Gamma(\omega) = \frac{1}{\hbar}(\mathbb{V}^S\eta_S(\omega)+\mathbb{V}^L\eta_L(\omega))$.

%------------------------------
\begin{figure}[t]
\begin{center}
\includegraphics[width=0.48\textwidth]{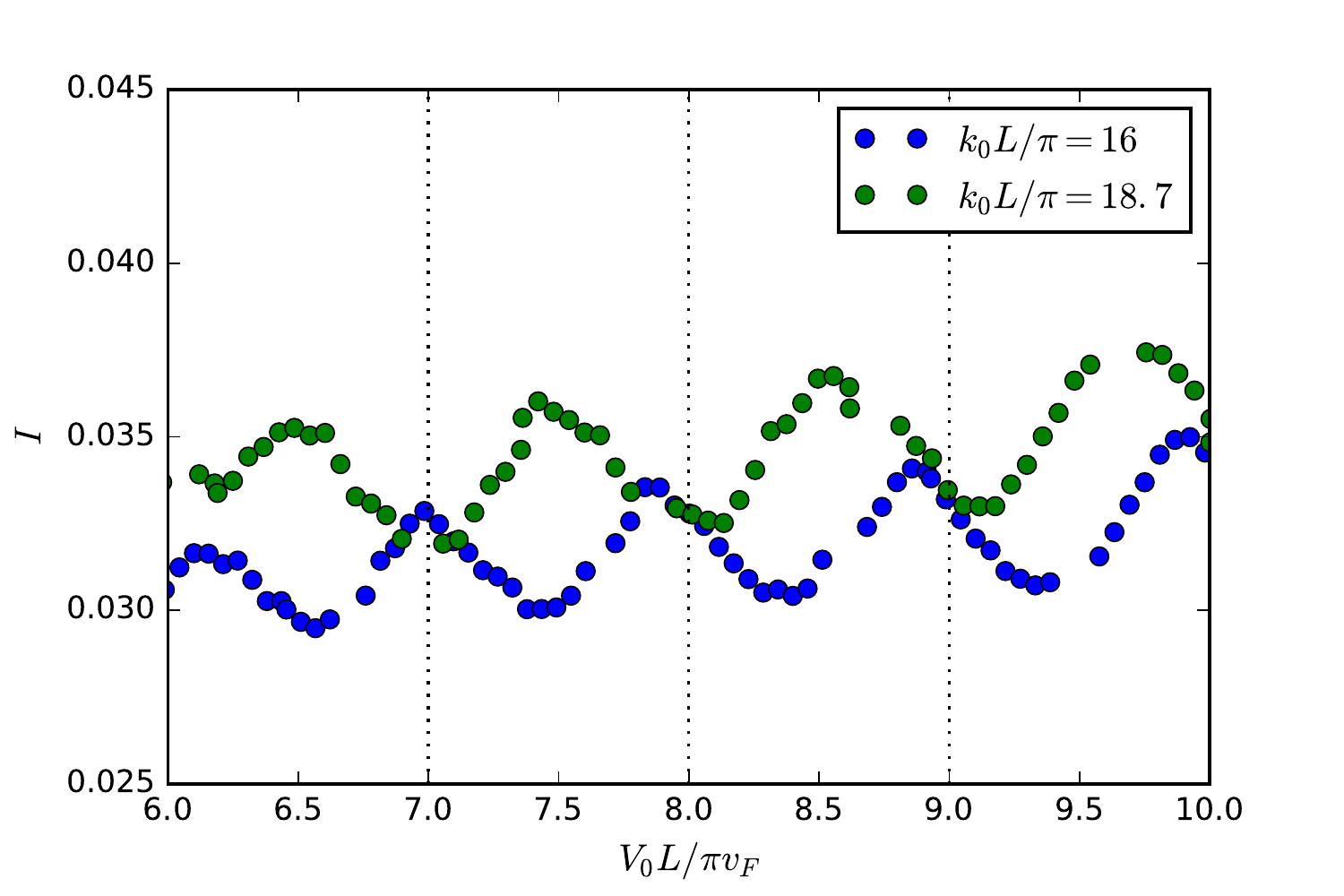} 
\includegraphics[width=0.48\textwidth]{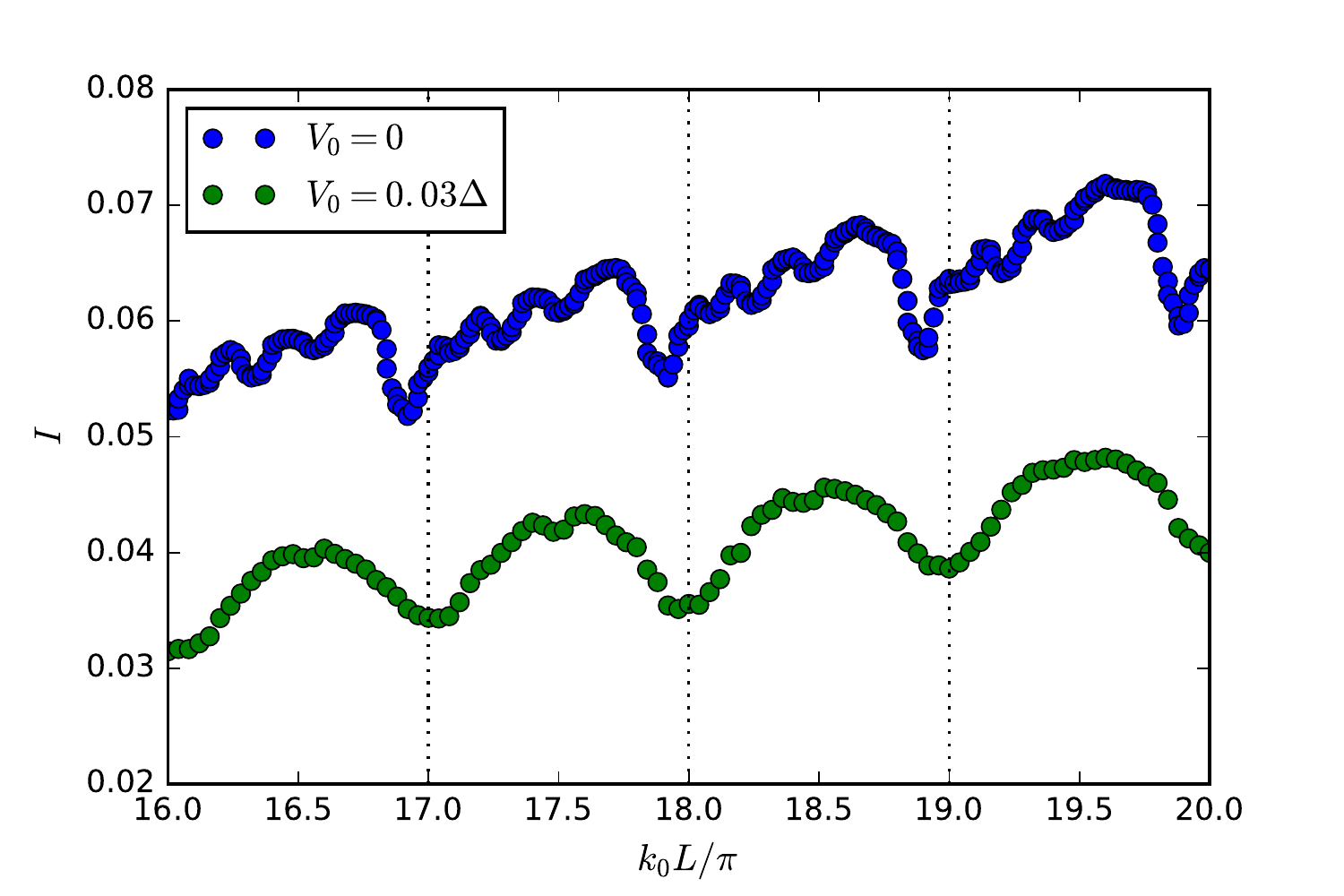} 
\end{center}
\caption{(a) Variation of the current as a function of the barrier height for an inversion symmetry broken WSM. The length of the Weyl semimetal is kept fixed at 100 sites. The values of the other parameters are $t_h=1$, $\mu_L=0.5\Delta$ and $\mu_R=0$.
 Here, $k_0= \pi/2 - {\rm sin}^{-1} (m/t_h)$ is kept fixed.  (b) The same as a function of the separation of Weyl nodes ($m$ is varied to change the separation of the Weyl nodes) in the Brillouin zone for fixed barrier height. The dotted lines indicate the periodicity expected from the ideas of the main text.}\label{fig:S2}
\end{figure}
%------------------------------

When the system is finite along the $z$ direction and periodic along $x,y$:
 \begin{align}
    \dot{N}_z = & \frac{i}{\hbar}[H,N_z] \nonumber  \\
    = & \frac{i}{\hbar}(- {\tilde t}+\lambda_z\tau)\big(\Psi_{z+1,\bar{\tau},\sigma}^{\dagger}(t)\Psi_{z,\tau,\sigma}(t) \nonumber \\ 
    & \hspace{2.2cm}  -\Psi_{z,\tau,\sigma}^{\dagger}(t)\Psi_{z+1,\bar{\tau},\sigma}(t)\big).
 \end{align}
Here we have used the explicit form of the Hamiltonian in the main text of the paper. So, the current along $z$ from a given site $z$ to $z+1$:
\begin{equation}
\begin{aligned}
  J_z(t)=\frac{ie}{\hbar}(-{\tilde t}+\lambda_z\tau)\Big(& \langle \Psi^{\dagger}_{z+1,\bar{\tau},\sigma}(t)\Psi_{z,\tau,\sigma}(t)\rangle \\  - & \langle\Psi_{z,\tau,\sigma}^{\dagger}(t)\Psi_{z+1,\bar{\tau},\sigma}(t)\rangle\Big) .
\end{aligned}
\end{equation}
Now, Fourier transforming the field operators, we have,
\begin{align}
&\langle \Psi_{z,i}^{\dagger}(t)\Psi_{z+1,j}(t)\rangle =  \int_{\omega, \omega'}\langle \Psi^{\dagger}_{z,i}(\omega)\Psi_{z+1,j}(\omega')\rangle e^{i(\omega-\omega')t},
\end{align}
with
\begin{align}
&\langle \Psi_{z,i}^{\dagger}(\omega)\Psi_{z+1,j}(\omega')\rangle \nonumber \\
&=\sum_{P,P'}\mathcal{G_W}_{z+1,I;jm}(\omega)\zeta^P_{ml}(\omega)\mathcal{G_W}_{Iz;li}^{\dagger}(\omega)\delta(\omega-\omega'). \nonumber
\end{align}
Here $\zeta^P_{ml}(\omega) = (\mathbb{V}^P_I\rho^P(\omega)\mathbb{V}^{P\dagger}_I)_{ml}$ where $\{I,P\}$ is either ${\{1,L\}}$ or ${\{N,SC\}}$ denoting either the normal or the superconducting lead respectively.

Putting everything back in, we can finally evaluate the current
\begin{equation}
  J_z(t) = e \rm{Im}\int d\omega Tr\Big[ \mathcal{A}\mathcal{G_W}_{z,I}(\omega)\sigma^P(\omega)\mathcal{G_W}^{\dagger}_{z+1,I}(\omega)\Big],
  \label{Current}
\end{equation}
where $\mathcal{A}_{31} = \mathcal{A}_{42} = -{\tilde t}+\lambda_z$ and $\mathcal{A}_{13} = \mathcal{A}_{24} = -{\tilde t}-\lambda_z$ and $\mathcal{A}_{ij} = 0$ otherwise. For the superconducting part, we obtained the Greens function by recursively solving for the surface of an $s$-wave superconductor. Also, we imposed the $flatband$ approximation for the normal lead. Hence, $\Sigma_L(\omega)=\mathbb{V}^L\mathcal{G}_L(\omega)\mathbb{V}^{L\dagger}=-i\pi\mathbb{V}^L\mathbb{V}^{L\dagger}=-i\pi\rho_L$. For this calculation, we have used $t^{SC} = t^L=0.25$. 
The values of the other parameters are given in the main text. 

\subsection{Inversion symmetry broken WSM}
The Hamiltonian used to describe an inversion symmetry broken WSM is
\begin{align}
  H^W =& \sum\limits_{\textbf{r}} \big(\Psi^\dagger_{\textbf{r}}(t_h\eta^zI_\tau\sigma^y)\Psi_{\textbf{r}+x} + \Psi^\dagger_{\textbf{r}}(t_h\eta^zI_\tau\sigma^y)\Psi_{\textbf{r}+y}\nonumber \\ & + \Psi^\dagger_{\textbf{r}}(t_h\eta^zI_\tau\sigma^y)\Psi_{\textbf{r}+z} + (m+2)\Psi^\dagger_{\textbf{r}}\eta^z\tau^y\sigma^y\Psi_{\textbf{r}} \big) \nonumber \\ & -\frac{1}{2}\sum\limits_{\langle\textbf{r}\textbf{r}'\rangle} \Psi^\dagger_{\textbf{r}}\eta^z\tau^y\sigma^y\Psi_{\textbf{r}'},
\end{align}
where $t_h$ is the hopping element inside the WSM and $m$ is the mass parameter. (For simplicity, we have combined spin-orbit couplings and hoppings and used $t_h$  to denote it). This model describes a normal insulator when $m>t_h$ and a Dirac semi-metal when $m=t_h$ with two nodes at $k_y=\pm\pi/2$. When $m<t_h$, each of the nodes split into two Weyl nodes forming a Weyl semi-metal with 4 nodes. The 4 Weyl nodes are located at $\pm k^+=\pm \sin^{-1}(m/t_h)$ and $\pm k^-=\pm(\pi - \sin^{-1}(m/t_h))$. Here, $k^{\pm}$ are defined in congruence with the main text.  Note that for this model,
$k^++k^-=\pi$ is fixed. The relevant inter-nodal distance is $k^+-k^- = k_0$. We keep the $x$ and $z$ directions periodic and the $y$ direction finite. Repeating the calculations for this setup, we end up with the same expression for the current (i.e, Eq.\ref{Current}) with $\mathcal{A}$ redefined such that $\mathcal{A}_{21} = \mathcal{A}_{43} = t_h$ and $\mathcal{A}_{12} = \mathcal{A}_{34} = -t_h$ and $\mathcal{A}_{ij} = 0$ otherwise. The results are summarized in Fig.~\ref{fig:S2}, and clearly, the two basic periodicities of the current as emphasized in the main text are seen here as well.

\end{document}